\documentclass[sigconf, authorversion,nonacm]{acmart}
\usepackage[frozencache,cachedir=.]{minted}
\usepackage{stfloats} 
\usepackage{booktabs}
\usepackage{soul}
\usepackage{tikz}
\usepackage{pifont}

\definecolor{bg}{HTML}{F0F0F0}

\newcommand\gc{Globus Compute}
\newcommand\name{CORRECT}

\usepackage{pgfplots}
\pgfplotsset{compat=1.18}
\usepackage{xcolor}

\usepackage[font=small,skip=4pt]{caption}

\AtBeginDocument{%
  }

\def\equationautorefname~#1\null{Eq.~(#1)\null}



\begin{document}


\title{Addressing Reproducibility Challenges in HPC with Continuous Integration}


\author{Valérie Hayot-Sasson}
\orcid{}
\affiliation{%
  \institution{University of Chicago}
  \institution{Argonne National Laboratory}
  \city{Chicago}
  \state{Illinois}
  \country{USA}
}
\email{vhayot@uchicago.edu}

\author{Nathaniel Hudson}
\orcid{}
\affiliation{%
  \institution{University of Chicago}
  \institution{Argonne National Laboratory}
  \city{Chicago}
  \state{Illinois}
  \country{USA}
}
\email{nhudson5@illinoistech.edu}

\author{Andr\'e Bauer}
\orcid{}
\affiliation{%
  \institution{Illinois Institute of Technology}
  \city{Chicago}
  \state{Illinois}
  \country{USA}
}
\email{abauer7@illinoistech.edu}

\author{Maxime Gonthier}
\orcid{}
\affiliation{%
  \institution{University of Chicago}
  \institution{Argonne National Laboratory}
  \city{Chicago}
  \state{Illinois}
  \country{USA}
}
\email{mgonthier@uchicago.edu}

\author{Ian Foster}
\orcid{}
\affiliation{%
  \institution{University of Chicago}
  \institution{Argonne National Laboratory}
  \city{Chicago}
  \state{Illinois}
  \country{USA}
}
\email{foster@uchicago.edu}

\author{Kyle Chard}
\orcid{}
\affiliation{%
  \institution{University of Chicago}
  \institution{Argonne National Laboratory}
  \city{Chicago}
  \state{Illinois}
  \country{USA}
}
\email{chard@uchicago.edu}

\begin{abstract}

The high-performance computing (HPC) community has adopted incentive structures to motivate reproducible research, with major conferences awarding badges to papers that meet reproducibility requirements. Yet, many papers do not meet such requirements. The uniqueness of HPC infrastructure and software, coupled with strict access requirements, may limit opportunities for reproducibility. In the absence of resource access, we believe that regular documented testing, through continuous integration (CI), coupled with complete provenance information, can be used as a substitute. Here, we argue that better HPC-compliant CI solutions will improve reproducibility of applications. We present a survey of reproducibility initiatives and describe the barriers to reproducibility in HPC. To address existing limitations, we present a GitHub Action, CORRECT, that enables secure execution of tests on remote HPC resources. We evaluate CORRECT's usability across three different types of HPC applications, demonstrating the effectiveness of using CORRECT for automating and documenting reproducibility evaluations.

\end{abstract}

\maketitle

\section{Introduction}



Modern computational and experimental methodologies have created new challenges in reproducing 
research outputs~\cite{sandve2013ten, hoefler2015scientific, papadopoulos2019methodological}. Reproducibility, defined as the ability for a different team to obtain the same measurements by following identical methodology \cite{acm_badges}, is often constrained by lack of access to code and data and insufficient documentation.
To promote reproducibility in science, initiatives such as \textit{Findable, Accessible, Interoperable, and Resuable}~(FAIR) principles~\cite{fair},  
reproducibility badges~\cite{badges}, and open science~\cite{opensci1, opensci2} 
have emerged. However, High Performance Computing~(HPC) infrastructure still poses a major challenge to reproducibility due to their typically specialized infrastructure, proprietary 
software, and restrictive security policies that prevent broad access to systems.

In recent years, there has been a push to make both code and data available; however, code and data alone are insufficient for reproducibility in HPC. Code may be difficult or impossible to compile and install, 
requiring specific versions of software dependencies and hardware. For many software issues, use of virtual machines and/or containerized environments, coupled with recipes for building these environments, may facilitate some level of reproducibility; 
however, it remains difficult to reproduce results on specialized hardware that may not be widely available.  
Consequently, open scientific data may not be meaningful if the software cannot be reliably run on HPC to reproduce the results. 


\begin{figure}[t]
    \centering
    \includegraphics[width=\linewidth]{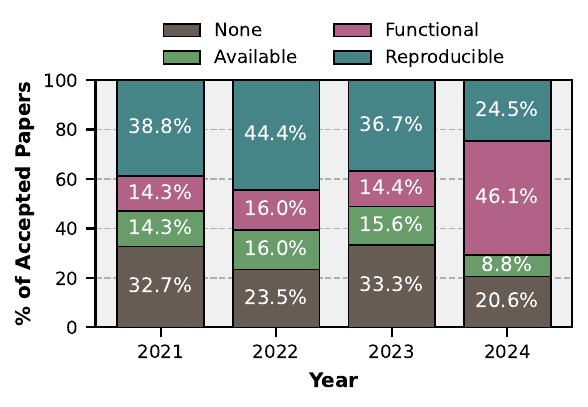}
    \caption{Reproducibility badges awarded by SC over time.}
    \label{fig:scbadges}
\end{figure}

Recent initiatives have sought to promote reproducibility in HPC, for example through reproducibility awards (e.g., SC's Best Reproducibility Advancement Award) and reproducibility badges, as well as 
organizations like IEEE and ACM launching reproducibility 
working groups and conferences focused on reproducibility.
While these advancements have incentivized the community to adopt reproducible practices, 
many HPC papers remain irreproducible (\autoref{fig:scbadges}). Moreover, the awarding of reproducibility badges is a largely manual process that is limited by both the hardware required and scale of the project. 


To ensure a minimum standard in code quality, conferences like CCGrid recommend \textit{Continuous Integration}~(CI) for automated software testing as part of their badge awarding process ~\cite{ccgrid_badges}.
Previous work~\cite{krafczyk2019scientific, jimenez2017popperci, sanz2022neuroci, fernandez2018continuous}, has demonstrated the effectiveness of incorporating CI practices into reproducibility evaluation, and coined the term \textit{Continuous Reproducibility}~\cite{fernandez2018continuous}.
CI is a standard software engineering practice to ensure that code changes do not introduce bugs (regression testing). 
Collaborative version control systems (e.g., GitHub~\cite{github}, GitLab~\cite{gitlab}, and Bitbucket~\cite{bitbucket}) provide mechanisms to define automated CI workflows, where tests are automatically run and results are published, and allow developers to display badges that represent the success of their test coverage. These systems promote a form of reproducibility by accounting for how the environment is configured, input arguments, and execution results, and can be tested across different environments, validating the observed performance across different infrastructures. However, they make the assumption that the code can be executed on arbitrary hardware (e.g., cloud-hosted VMs) or that the user can set up their own runner for CI. Neither approach is suitable for HPC.



To address the limitations of current methods for reproducibility, and more specifically, CI in HPC, we propose a GitHub Action that builds on \textit{\gc{}}~\cite{chard2020funcx}, a Function-as-a-Service (FaaS) platform that can execute functions across available computing infrastructure (e.g., HPC, cloud, workstations). Using this approach, users can automate the execution of reproducibility-focused unit test cases on specialized infrastructure through a typical GitHub flow.
The contributions of this paper are as follows:
\begin{itemize}
    \item A review of current technologies and initiatives that enable reproducibility and CI in HPC.
    \item CORRECT, a GitHub Action that enables automated reproducibility evaluations across HPC and other remote sites.
    \item An evaluation of \name{} using three different scientific applications for HPC.
\end{itemize}


\section{Reproducibility in Science}
Criteria that promote the reproducibility of scientific code include:
\textit{(i)}~code and data being published and publicly available, 
\textit{(ii)}~adequate documentation, and 
\textit{(iii)}~details on software dependencies and infrastructure. 
Significant efforts have been made to represent and publish structured collections of digital resources related to a scientific investigation, or \textit{Research Objects}~(RO)~\cite{soilandreyes22rocrate, dupre2022beyond, lohrey2009integrated, edmunds2017experiences, bechhofer2010research}. 
Several platforms, such as Whole Tale~\cite{chard2020toward} and Code Ocean~\cite{cheifet2021promoting}, have been developed to
promote the implementation and sharing reproducible research. 
%

Whole Tale is a web platform for generating re-executable research objects called ``Tales''. These objects contain details on code, data, and compute environment, as well as any other documentation. 
Users can access Whole Tale to conduct, share, publish, and re-execute experiments. 
Whole Tale is implemented as three main components: a web-based front-end, a management web service, and an execution environment. 
The web service is built on Girder---an open source data management platform developed by Kitware---and provides APIs to manage users, jobs, tales, and data. The web interface exposes access to crate and reproduce Tales using various supported interfaces, such as Juypter. The execution environment, in which Tales are created and run, relies on Docker~\cite{docker} and has been deployed at both TACC and NCSA. The Whole Tale team investigated the challenges of running Tales on HPC systems and proposed several models that could be used~\cite{chard2020toward}; however, this capability has not been implemented in Whole Tale. 

Code Ocean is a cloud-based platform designed to support reproducibility and collaboration in computational research. It enables creation and sharing of  \textit{capsules}---self-contained, shareable, and executable environments that bundle code, data, metadata, and a fully-specified compute environment. Capsules can be run directly in the browser, enabling easy validation and re-execution without any manual setup. Code Ocean 
supports popular programming languages (e.g., Java, Python, C/C++, R, and MATLAB) and execution environments defined via Dockerfiles and built on  predefined base images. It also integrates version control and assigns persistent identifiers (DOIs) to published capsules. Widely used in the publication process by journals and conferences, Code Ocean promotes reproducibility during the peer review process. While the platform supports container-based workflows, it is primarily focused on small- to medium-scale computational tasks and does not natively support HPC deployments.

RO frameworks, such as RO-Crate~\cite{soilandreyes22rocrate}, help facilitate reproducibility of scientific computing by providing a FAIR method for packaging data and their associated metadata, including, for example, computational workflows. However, to the best of our knowledge, RO-Crate does not have the ability to integrate HPC platforms. 

The remainder of this section discusses the various components and frameworks integral to facilitating reproducibility of scientific research. 





\subsection{Containerization and Environment Management}

Achieving computational reproducibility in science requires capturing the entire software environment/stack. One possible solution is with virtualized environments. 
The traditional approach, virtual machines (VMs),  
virtualize the entire hardware stack, including the operating system, and  therefore isolate individual applications by running them in a virtualized environment 
on a given machine; however, they can be resource-intensive. An alternative approach to virtualization is using containers. Containers virtualize at the operating system level and are therefore often more efficient and smaller, but provide less isolation, than VMs. \textit{Docker}~\cite{docker}, the most popular container platform, pioneered packaging applications with all dependencies into portable containers. However, its reliance on a privileged daemon 
is rarely supported in shared HPC environments. As a result, solutions like \textit{Singularity}~\cite{kurtzer2017singularity} (now \textit{Apptainer}), \textit{Shifter}~\cite{shifter}, and \textit{Charliecloud}~\cite{priedhorsky2017charliecloud} were developed. These container platforms are specifically designed for scientific and HPC workflows, allowing researchers to run containerized environments without needing root privileges, compiling with HPC security models and resource managers. 

Complementary to containerization are robust package and environment managers. For instance, \textit{Conda}~\cite{conda} is a package and environment manager primarily used for Python and R software, while \textit{Spack}~\cite{spack} offers fine-grained control for building HPC software from source.
GNU Guix~\cite{vallet2022toward} targets reproducibility via declarative package management emphasizing bit-reproducible builds and transactional control. It enables instantiation of reproducible environments and can directly generate container images with reproducibly composed contents.

Ideally, the outcome of a scientific computation would be solely a function of the defined hardware and software environment. However, the inherent complexity of modern computer systems introduces significant variability, stemming from factors unrelated to the packaged code itself. Subtle fluctuations in runtime execution---such as the precise timing of thread scheduling, dynamic compilation effects, physical data layout on storage, network contention, power management adjustments, or even temperature variations influencing hardware behavior---can potentially lead to minor differences in results between identical runs.

Despite inherent systemic variability, this layered approach, combining meticulous package management with robust containerization, provides a comprehensive framework for encapsulating and sharing complete computational ecosystems, facilitating the verification and reproduction of scientific findings across diverse computing infrastructures.

\subsection{Distributed Deployment}
\subsubsection{Scientific Workflows}

Modern scientific applications often involve complex, multi-step computational analyses, where managing dependencies and ensuring reproducibility is paramount. Scientific workflow management systems were designed to support such applications. 
Tools such as \textit{Snakemake}~\cite{molder2021sustainable}, \textit{Nextflow}~\cite{nextflow}, \textit{Pegasus}~\cite{deelman2015pegasus}, \textit{Parsl}~\cite{babuji2019parsl}, and \textit{Toil}~\cite{vivian2017toil} enable programmatic definition of the entire workflow structure---the sequence of tasks, their inputs/outputs, and their interdependencies. By providing these precise, shareable blueprints, these systems enhance reproducibility. Workflow management systems often integrate seamlessly with container technologies, providing a basis for capturing execution environments.

As the workflow runs, a detailed log is created that captures exactly what was run, how, and when. This provenance includes, among other information, the specific versions of the workflow script and software tools used, the exact parameters supplied for each step, timestamps for task execution, resource usage information, and logs generated by each task. These records enable researchers to verify findings and reliably rerun the analysis under the same conditions. 

\subsubsection{Multi-Site Deployments}

Due to the nature of scientific computing, diverse and distributed resources are often required, which presents challenges as workflows may span multiple administrative domains and infrastructures. To address this, various frameworks and platforms have been developed to support multi-site deployment and execution. 
For example, \textit{\gc{}}~\cite{chard2020funcx} is a hybrid cloud-edge platform that enables researchers to execute Python functions across a federated network of registered endpoints---simplifying use of heterogeneous systems.
Similarly, middleware platforms (e.g., \textit{Apache Airavata}~\cite{airavata}, \textit{Tapis}~\cite{tapis}) offer services to manage computations across distributed resources. These systems provide APIs that abstract site-specific details, often integrating federated identity mechanisms for authentication. 
Collectively, these frameworks support execution on a wide range of target environments, including university research clusters, national supercomputing facilities (such as ACCESS in the US or PRACE in Europe), commercial cloud platforms (e.g., AWS, Azure, GCP), and institutional or laboratory servers.

\section{Reproducibility in HPC} 

Reproducibility in HPC has largely been incentivized through reproducibility badges and awards, requiring a formal review process to ensure reproducibility of research artifacts. The badge process itself may recommend the use of CI as an aid to reproducibility, but it is often not required. Reasons for this include HPC research requiring access to specialized hardware, HPC-optimized software, or significant computing resources. HPC sites make the implementation of CI frameworks for HPC challenging, due to security and access restrictions, preventing remote automated invocation of computing jobs. In this section we outline the reproducibility badge awarding process as well as discuss the available CI frameworks for HPC.


\subsection{Reproducibility Badges}

Reproducibility badges are awarded after a formal review process. They serve as proof that independent parties were able to reproduce core results. Systems conferences have adopted this process as a means to evaluate and promote reproducibility in HPC. 

\subsubsection{Available badges} 

Reproducibility badges at SC and CCGrid are  organized in three levels.
Higher-level badges imply that lower-level requirements have been met.

The first level is typically called "Artifacts Available" or "Open Research Objects".   
This badge indicates that the software and input data used by the authors are stored in a permanent public repository under an open license.
Documentation must be sufficient for reviewers to understand the core functionality of the software, and any input data should be accompanied by some context and descriptions.
Repositories such as Zenodo, FigShare, Dryad, and Software Heritage are commonly accepted as permanent storage.

The second level is usually called ``Research Objects Reviewed'' or ``Artifacts Evaluated''.
This badge confirms that the reviewers have installed the software and verified its core functionality with a small experiment.
At a minimum, the software should provide manual instructions to check its expected behavior (such as compiling and executing a test run).
Ideally, it also includes an automated test suite using CI systems.

The last level, ``Results Reproduced'' or ``Results Replicated'', is the most challenging badge and is awarded when reviewers successfully reproduce the paper's key computational results.
The aim is to validate the central claims and findings of the paper rather than to achieve identical results, particularly when hardware-specific differences might occur.

\subsubsection{Review methodology}
Reproducible experiments are generally referred to as "artifacts" when evaluating them.
For the SC conference, authors are required to describe such artifacts in two ways: an \textit{Artifact Description}~(AD) and an \textit{Artifact Evaluation}~(AE).
The AD presents details about the artifacts in a machine-agnostic way. It lists the main contributions of the paper and specifies which experiments should be reproduced.
It describes the experimental results by referring to the general trends observed, without going into the specific values of the results.
The AE provides more detailed instructions on how to install the code, offers usage guidelines, and describes the experimental workflow; simply asking users to run a script is insufficient. The AE illustrates how to run the AD on a specific machine with a specific compiler, providing version information, required packages, and the exact steps needed to reproduce the results.

From these two documents, reviewers must evaluate the reproducibility of the artifacts. 
Reviewers are expected to follow the steps outlined in the AE. If they encounter installation problems, they are encouraged to contact the authors for assistance rather than 
troubleshooting on their own.
Reviewers are usually given a limited amount of time, typically about eight hours or one business day, to reproduce the experiments. 
In the report, reviewers describe the time required, 
machines used, problems encountered such as implicit assumptions, missing documentation, versioning issues, data accessibility or hardware-specific issues, or missing environment variables, and whether it matches the contribution described in the AD.

\section{Continuous Integration }

\textit{Continuous Integration}~(CI) is a software development practice in which code changes are frequently integrated and validated through automated testing. CI provides many benefits, such as regularly verifying the proper functioning of the integrated code. CI is important in the context of reproducibility, as it can track how code changes impact behavior in different environments. 

There are many CI services available, such as GitHub Actions~\cite{github_actions}, GitLab CI/CD~\cite{gitlab_ci}, Travis-CI~\cite{travis_ci} and Jenkins~\cite{jenkins}. Each of these platforms rely on a workflow description file, commonly written in YAML or JSON, to define the steps required to download, build, and test the code. CI services are either fully integrated with, or provide plugins to work with, popular cloud-hosted version control solutions such as GitHub and GitLab. 

Being both based on git versioning, GitHub and GitLab provide similar capabilities. However, they differ in terms of cost model and 
functionality. For instance, GitLab is entirely open-source, enabling enterprises to deploy their own local GitLab services, ensuring the security of their enterprise needs. In contrast, GitHub is closed-source, but is rich in features that promote large-scale collaboration.  As both GitLab and GitHub are commonly used for hosting the versioning on open-source research applications, we will discuss their integrated CI solutions in more detail.

\subsection{GitHub Actions}
GitHub Actions workflows consist of three parts: 
\textit{(i)}~\texttt{events} that trigger the workflow, 
\textit{(ii)}~\texttt{jobs} composed of \texttt{steps} that are invoked by the workflow, and 
\textit{(iii)}~\texttt{runners} on which the jobs  execute.

GitHub Actions supports a variety of events that can trigger workflows. These events include scheduled events, code changes to the repository, and changes that occur outside the repository. 

Jobs within a GitHub Actions workflow may interact with services that require authentication (e.g., DockerHub). In order to bypass manual authentication, application \texttt{secrets} can be stored in GitHub to be used in workflows. These secrets can be stored in the organization, repository, or in an environment for that repository. In contrast to other types of secrets, environment secrets allow repository administrators to specify access permissions for the secret. These permissions include adding reviewers to authorize use of environment, specifying a wait timer for the environment, and  defining which branches can access the environment. Secrets cannot be specified per user, preventing finer-grained access control.


Steps within GitHub Actions jobs may call third-party actions published in the GitHub Action Marketplace~\cite{github_marketplace} or custom actions defined by the workflow. Third-party actions simplify the workflow description and eliminate the need for duplicate code. 


GitHub Actions workflows are executed on machines referred to as \texttt{Runners}. GitHub hosts their own set of runners on Microsoft Azure that can be used by workflows. These runners are virtual machines pre-configured with Linux (Ubuntu), Windows, or MacOS and are equipped with x64, Intel, or ARM64 architectures. Larger runners are also available for paid plans, such as GitHub Team and GitHub Enterprise Cloud, and include larger VM instances and Tesla T4 GPUs. While these runners provide extra flexibility, they lack the flexibility for users to specify the underlying hardware. 
GitHub Actions provides users with the ability to use their own self-hosted runners deployed on their own resources.




\subsection{GitLab}
Like GitHub, 
GitLab CI/CD pipelines are described in YAML, define different stages (e.g., \texttt{test}, \texttt{build}, \texttt{deploy}) and the jobs which run within them. Jobs are executed in runners that are either cloud- or self-hosted and can make use of \texttt{components}, GitLab's equivalent to \texttt{actions}, to reduce code duplication or refer to custom code.

Pipelines in GitLab can be triggered manually or automated using scheduled or pipeline triggers. Scheduled triggers allow users to automate the pipelines using a predefined time interval, whereas pipeline triggers are GitLab authentication tokens that can be used within REST API requests to invoke a pipeline. Unlike GitHub, the triggers themselves are not pre-defined. Users can manually trigger pipelines by directly invoking the API call with the tokens, specifying a rule-based policy in the CI/CD pipeline, or via another pipeline's web-hook. 

To authenticate to external services, GitLab supports secrets. GitLab provides native integration with several secret management providers, including Azure Key Vault~\cite{azure_key_vault}, HashiCorp~\cite{hashicorp}, and Google Cloud Secret Manager~\cite{google_secret_manager}. Using secret manager providers is the most secure way to store and use secrets in a pipeline, although GitLab also provides \texttt{variables} that function similarly to GitHub secrets, storing the secret within the project, group, or instance. Users with access to the settings have access to view their unmasked values, but they can be hidden or masked. Variables can also be protected, running only on specified protected branches.

GitLab Components are small units of code that can create a larger pipeline. Components can be custom-made, located in their own unique repository and reused across projects, or be a third-party tool published in the CI/CD catalog~\cite{gitlab_ci_catalog}. Use of third-party tools is subject to the same vulnerabilities as GitHub Actions.

GitLab provides several hosted VM runners, including Linux (both ARM64 and AMD64) as well as GPU runners (NVIDIA Tesla T4), Windows, and macOS (M1 and M2). Only the Linux and Windows runners are available in the free tier, with other runners reserved for Premium and Unlimited tiers. As with GitHub-hosted runners, it is impossible to select specific infrastructure, requiring the user to rely on self-managed runners to test on specific hardware components.

\subsection{CI for scientific computing}

While reproducibility of scientific applications has recently been measured using CI~\cite{krafczyk2019scientific, jimenez2017popperci, sanz2022neuroci}, CI has been implemented in various large scientific software projects. These projects all have similar characteristics, as outlined in \autoref{tab:ci-science}.
To help with evaluating software reproducibility, CI frameworks provide extensive logging, visualization, and monitoring of runs. Here we describe various CI solutions implemented for different scientific use cases spanning high energy physics, molecular chemistry, neuroscience, and signal processing. These solutions are summarized in  \autoref{tab:ci-science-frameworks}.
\begin{table}[]
    \centering
    \small
    \caption{Science application features important for CI.}
    \label{tab:ci-science}
    \begin{tabular}{p{2.3cm}p{5.4cm}}
    \toprule
         \textbf{Characteristic} & \textbf{Description} \\
    \midrule
        Collaboration & Scientific software consists of multilayered code \\
    \midrule
        Computational requirements & Applications may  process large volumes of data, require substantial amounts of memory, and take a long time to test \\
    \midrule
        Visualization, Monitoring, Logging & It is important to be able to monitor execution, visualize changes, and access historical information \\
    \midrule
        Reproducibility & Performance and accurate downstream results is important \\
    \bottomrule
    \end{tabular}
\end{table}

\begin{table*}[t]
    \centering
    \small
    \caption{Comparison of different CI framework usage in scientific applications}
    \label{tab:ci-science-frameworks}
    \begin{tabular}{r|cccc}
    \toprule
        \multicolumn{1}{}{}
        & \textbf{GNSS-SDR} & \textbf{ATLAS} & \textbf{AMBER} & \textbf{NeuroCI}  \\
    \midrule
        CI framework & GitLab & Jenkins & Cruise Control & CircleCI \\  
        Compute resource & Cloud & Internal HPC cluster & Workstation & Distributed HPC clusters \\
        Objective & Reproducibility & CI & CI & Reproducibility \\
        Visualization & Stored artificats & Monitoring Dashboard & GNUPlot performance plots & Scatter/Distribution plots \\
    \bottomrule
    \end{tabular}
\end{table*}

\subsubsection{ATLAS}
ATLAS is a large collaborative project that implements comprehensive testing with CI~\cite{atlas2025software,elmsheuser2017roadmap}. 
ATLAS CI polls the GitLab server to determine if code changes have been proposed in merge requests. Testing and deployment is also performed on a nightly basis.
CI is facilitated via the Jenkins framework installed on a 1400 core build-farm for execution. This environment includes a mixture of real hardware and virtual nodes. The testing suite consists of unit tests and short integration tests. ATLAS prioritizes fast delivery of results, relying on ``operation intelligence'' techniques (e.g., incremental testing, selective compilation) for efficient resource use and minimal execution time. To monitor CI, ATLAS provides a monitoring system that includes details on installation, build, and test results; monitoring information is available via the BigPanDA web application.

\subsubsection{AMBER}
Assisted Model Building with Energy Refinement (AMBER) is a popular molecular chemistry project that is developed by a team that spans multiple research institutions~\cite{betz2013implementing}. The code itself  supports many different architectures, GPUs, and major compilers. As this code is primarily maintained by researchers, maintainers are focused on the components related to their research, thus it is imperative that the code base is regularly tested against different conditions. 

The build process is automated using GNU Make and CI is provided by 
CruiseControl~\cite{cruisecontrol}.
AMBER attempts to promote short test cycles such that developers can obtain timely feedback.
To speed up CI time, they host the CI server on a dedicated machine that can support the building and testing of multiple targets in parallel. They also apply staged builds to ensure timely feedback on the correctness of the code. 
Virtualization is used to enable testing on different software environments.

After testing, a script is executed to collect all the ``diff'' files produced by failed tests. 
Failures are displayed on the web interface. Since performance impacts of code changes are also relevant, timestamps are created before and after execution and a graph is generated 
and displayed on the web interface. Build artifacts are saved and made available via the dashboard for persistent access to historic information of the runs.

\subsubsection{NeuroCI}
NeuroCI is a CI framework developed to capture the variability that may occur when performing neuroimaging analyses across different processing pipelines and datasets~\cite{sanz2022neuroci}. Due to both the large memory requirements and large amounts of data used in neuroimaging applications, cloud-hosted CI solutions are not favorable. 
NeuroCI addresses the challenges of processing neuroimaging pipelines through CBRAIN~\cite{sherif2014cbrain} for distributed computing and DataLad~\cite{wagner2022fairly} for data management. To automate the CI execution, NeuroCI uses CircleCI. 

Workflows in NeuroCI are described in JSON and specify the command to be executed, filepaths of parameters, and any other input data required.
Since users must authenticate with CBRAIN, access secrets are stored in CircleCI. When repository code is updated, or the periodic interval is met, the CI is executed. This process involves using a base Docker container to build repository code and prepare tasks for launch.
To execute tasks on remote infrastructure, a process known as the \textit{pipeline manager} submits the tasks to the CBRAIN API.

NeuroCI provides additional features relevant to reproducibility. For example, all task provenance data is gathered and stored within a task provenance cache file. The task provenance cache file stores CBRAIN IDs pointing to the location of the tasks and files within CBRAIN. These generated cache files are exported as artifacts within CircleCI and are made available through an API. Additionally, NeuroCI provides tools that facilitate result visualization, such as
distribution 
histograms and combined scatter/distribution plots for each pipeline/dataset combination. These plots are hosted on a Flask server deployed on the HPC infrastructure and the URL is shared via the CircleCI interface during execution. Users may also opt to generate their own visualizations from the data by modifying a Python module within NeuroCI.


\subsubsection{GNSS-SDR for Signal Processing Research}
Fernandez-Prades et al. explore the notion of \textit{Continuous Reproducibility}~(CR) in signal processing for global navigation satellite systems~(GNSS) \cite{fernandez2018continuous}. 
They implement CR as an extension to CI, triggered upon successful completion of CI, similar to \textit{continuous deployment} and \textit{continuous delivery}. 
CR 
integrates external data and (experimental) software execution to ensure that the final results are reproducible as updates to the software occur.
Their research explores many  common issues related to reproducibility, specifically as it relates to workload, system (hardware and software), and results, as well as GNSS-specific challenges related to systems and streaming data.
They present 
a prototype implementation of a CR solution for GNSS research. This implementation uses standard software for different pieces of the overall CR setup, including Git, GitLab, Zenodo-provided DOI, compilers (namely, GCC and LLVM/Clang), and Docker for containerization of software.
While GNSS-SDR is a great example of reproducibility in action, it remains application-specific. Furthermore, its design does not explicitly consider the needs of HPC-dependent scientific tasks.

\subsection{CI platforms for HPC}
While there is a strong need for CI in HPC projects, efforts to bring CI to HPC environments remain limited. The reason for this is due to several factors~\cite{gamblin2022overcoming}: availability of HPC micro-architectures in cloud-based CI, security, and administrative and political challenges (e.g., HPC systems are designed to run large production codes). Nevertheless, a few solutions have been developed to bring CI to HPC platforms adhering to the characteristics outlined in \autoref{tab:ci-hpc}. 
Here we discuss solutions that have been implemented at the Department of Energy (DOE), Texas Advanced Computing Center (TACC), Ohio Supercomputing Center (OSC), RMACC's Summit Supercomputer and at Stanford's High Performance Computing Center (HPCC). A summary of these systems is presented in \autoref{tab:ci_specs}.


\begin{table}[t]
    \centering
    \small
    \caption{Characteristics important for CI of HPC software}
    \label{tab:ci-hpc}
    \begin{tabular}{p{2cm}p{5.7cm}}
    \toprule
         \textbf{Characteristics} & \textbf{Description} \\
    \midrule
        Collaborative & HPC software is developed by many research groups with access to different infrastructure.
        \\ 
    \midrule
        Secure & User code executing on HPC should not gain elevated privileges and must be linked to the appropriate user account. 
        \\
    \midrule
        Lightweight & CI should be mindful of resource use.
        \\
    \bottomrule    
    \end{tabular}
\end{table}

\begin{table*}[t]
    \centering
    \caption{HPC CI frameworks  feature comparison}
    \label{tab:ci_specs}
    \resizebox{0.8\linewidth}{!}{
        \begin{tabular}{
            r
            ll
            cl
        }
        \toprule
            & & & \textbf{Site-Specific} &
        \\
            \textbf{Framework} & 
            \textbf{CI Platform} & 
            \textbf{Authentication} & 
            \textbf{Execution} &
            \textbf{Containerization}
        \\
        \midrule
            Jacamar CI  
            & GitLab
            & Site-Specific Auth.
            & Yes
            & Apptainer, Podman, CharlieCloud
        \\
            TACC 
            & GitHub
            & Tapis Security Kernel
            & No
            & Singularity
        \\
            RMACC Summit 
            & Jenkins
            & Site-Specific Auth.
            & Yes
            & Singularity
        \\
            OSC 
            & Reframe
            & Site-Specific Auth.
            & Yes
            & None
        \\
            Stanford HPCC 
            & GitLab
            & Site-Specific Auth.
            & Yes
            & Unknown
        \\
        \bottomrule
        \end{tabular}
    }
\end{table*}

\subsubsection{Jacamar CI}

Jacamar CI~\cite{adamson2024creating} was developed in response to the need for CI on specialized HPC infrastructure in the DOE Exascale Computing Project (ECP).  Jacamar CI integrates site-specific security requirements with GitLab's CI capabilities to provide a secure solution to run CI on HPC. The layer in which Jacamar CI operates occurs between the GitLab runners and the user's environment. 
Using this approach, Jacamar CI can ensure that the two principal security requirements for HPC are satisfied: 
\textit{(i)}~identity used to run the code matches the user who intended to launch it and 
\textit{(ii)}~users and/or processes launched by the CI cannot access or modify files or aspects of the system beyond their permission. Site-specific security requirements are also enforced. Jacamar CI ensures that the HPC identity used to run the code matches that of the user who invoked it through the use of a JSON Web Token that is generated and signed by the GitLab server. Additionally, Jacamar CI restricts the permissions of the CI job to match those of the user.

Runner processes run directly on the HPC site's login node, pulling the CI code directly from the repository. There are two modes for the runner processes: 
\textit{(i)}~shared runner and 
\textit{(ii)}~many-individual-runners. 
With the latter, each project runs its own runner, whereas, with a shared runner, there is a single runner deployed on the login node for all projects. Using a shared runner reduces administrative burden on HPC systems. 
When executing the runner job, Jacamar CI leverages the underlying HPC scheduler or executes the code directly within the user's Bash terminal. Post execution, job results are captured and returned to the user via the GitLab interface for review. 

As many HPC software repositories have collaborators that are external to the HPC infrastructure, solutions that do not restrict collaboration are necessary. Jacamar CI attempts to balance collaboration and security of the underlying infrastructure, by mirroring parts of the code that require testing on HPC infrastructure within the internally-hosted GitLab server. Nevertheless, the number of mirrors would still need to match the number of HPC sites that need to be used for testing.

\subsubsection{CI with Tapis}
As part of their development process, TACC wanted to integrate CI for integration testing~\cite{pachev2022continuous}. There were two main challenges that needed to be solved by their solution: 1) two-factor authentication impedes automation, and 2) the CI system could not control job execution using the batch scheduler. To tackle these challenges, Tapis~\cite{stubbs2021tapis} was used as the intermediary between GitHub Actions and  TACC's HPC clusters. Tapis contains a security subsystem, known as the Tapis Security Kernel, that integrates with any identity/authentication manager to facilitate authentication on TACC resources. Additionally, Tapis tackles the second challenge by providing the Tapis Jobs API to execute jobs on remote interfaces.

Each CI job needs to be associated with a Tapis application.
The application scripts can be used to pull Docker images and convert them to Singularity, alleviating the need to update the application regularly. The tests can then be executed directly within the latest Singularity image.

Rather than using GitHub Actions' default runners that run on the cloud, Tapis CI uses a self-hosted runner deployed on Jetstream. The self-hosted runner provides better control of the environment and guarantees reduced execution time by pre-configuring certain settings. To verify whether a job has been completed, the runner communicates with the Tapis webhook, avoiding the need to poll Tapis directly. Metadata from runs and outputs are then committed back to the repository.

\subsubsection{CI on RMACC's Summit}
With the growing need to improve the software development cycle, both for system engineers and cluster researchers, RMACC's Summit adopted a framework for CI~\cite{sampedro2018continuous}. The CI platform consists of Jenkins services deployed locally via Docker compose. Docker was selected for the deployment of the infrastructure, due to its rich set of services that could facilitate deployment at other sites. 

CI of projects are automated via a Jenkins job that polls repositories for changes in the code. As security would need to be more stringent for users, Singularity was adopted for the building of the software environment. All code repositories therefore contain a Singularity recipe file alongside the source code. When a CI job is triggered, Jenkins will build the Singularity image from the recipe, run the test cases, and publish the images to a self-hosted Singularity Registry.


\subsubsection{CI at OSC}
OSC implements CI workflows  to test and deploy new software environments~\cite{khuvis2019continuous}. Unlike many other frameworks for CI in HPC, OSC does not use GitLab or GitHub CI, due to difficulties with submitting batch jobs on their environment. Instead, an install script is executed to build and generate a module file for the target software library and push the module file to a GitLab repository. A webhook is used to detect any commit changes to the repository and trigger the CI. The ReFrame framework~\cite{reframe} is then used to execute the tests. To determine whether the tests passed or failed, a cron job executes daily that retrieves the outputs of the executions, sends an email to administrators to notify of any failures, and updates the module's repository README with the status of the tests.

Due to the use case for the CI, most of the steps are executed by administrators. Only ReFrame tests are executed with user-level permissions when evaluating user-facing software. This should not preclude the ability to expand this to user software, but as a result it is restricted to single-site use, relying on system authentication and an internal GitLab.


\subsubsection{CI at Stanford HPCC}
The code developed at Stanford HPCC, such as Hypersonics Task-based Research (HTR), must be evaluated against different target HPC centers \cite{melone2023verifying}. In the case of HTR, HPCC is used as a testbed environment for Lawrence Livermore National Laboratory (LLNL). LLNL provides an instance of Jacamar-CI for CI, whereas HPCC offers a scaled-down version of Jacamar-CI. This scaled-down implementation includes a self-hosted GitLab runner service deployed on an unprivileged user account. The runner listens to changes on the publicly-hosted GitLab and submits jobs to SLURM.

\section{Continuous Reproducibility for HPC}

Reproducibility of HPC software can be limited by the lack of access to the necessary resources. However, with sufficient accounting (e.g., previous execution runs and their results, system provenance, source code) and automated periodic reexecution demonstrating result validity, it is possible to evaluate reproducibility without direct access to the infrastructure. While CI can both automate reexecution of code and gather details relating to the system's software environment, there remain various challenges associated with implementing CI for HPC infrastructure. Although several solutions exist, they are often limited by the constraints imposed on services executing on HPC. We sought to develop a set of guidelines and build an easy-to-deploy and easy-to-use prototype that would address these limitations with two primary goals: 
\textit{(i)}~enable multi-site execution
and
\textit{(ii)}~satisfy basic security demands of HPC infrastructure.

To meet these requirements, we build upon existing software used by scientific and HPC communities for multi-site execution, security, and code versioning/CI. The two main tools used in our solution are: Globus Compute and GitHub Actions.

\subsection{Globus Compute} 

\gc{} is a hybrid FaaS platform that decouples function registration and management from function execution on a federated ecosystem of \gc{} endpoints. The cloud service provides a single contact point via which functions can be registered and submitted for execution. An executed function is called a ``task''. The cloud service manages the execution of one or more tasks on a selected endpoint. When a task completes, the endpoint returns the result, or exception, to the cloud service for users to later retrieve.

\gc{} endpoints can be deployed in single-user mode in user space or in multi-user mode as a privileged service~\cite{ananthankrishnan24compute}. In the latter case, the multi-user endpoint (MEP) forks a user endpoint (UEP) process in user space for the requesting user. MEPs use the same identity mapping process as used by Globus Connect Server~\cite{chard14efficient}.  
In both cases, endpoints can be configured to provision resources from different types of resources, such as HPC schedulers or Kubernetes clusters.  Endpoints use Parsl to dynamically provision resources, deploy a pilot job model, and manage the execution of tasks on those resources, optionally in a container.

\gc{} provides an ideal substrate on which to build our CI environment as it enables secure and reliable execution on arbitrary remote computing endpoints. Thus, CI can be configured to run tests on \textit{any} remote device by installing the \gc{} endpoint software. Endpoints implement a complex security model in which they communicate with the cloud service via secured, outbound connections. Pairing between users and endpoints is established via the OAuth-based Globus Auth platform.  Further, MEPs can be configured with different types of high assurance policies, for example, requiring specific identity providers, enforcing sessions, and restricting the functions that can be executed. MEPs are configured with templates that define what resources can be used by UEPs and administrators can audit logs of all tasks that have been executed.


\subsection{Addressing security challenges}

We rely on a combination of security features available in both \gc{} and GitHub Actions to meet the security needs of HPC applications. While GitHub Actions does not provide the ability to map secrets to users, ensuring that only the user who triggered the flow can invoke the flow, GitHub does provide environment secrets. These secrets are created by repository administrators and enable the ability to force review on trigger. Using environment secrets, CI workflows will not be executed until they are approved by the environment reviewer. This ensures that the person authorizing the execution maps to a user at the site at which the code is executed. Additionally, this gives the reviewer the opportunity to examine code changes and ensure that malicious code will not be executed at the site. Using environments, it is strongly suggested that there is only one reviewer per environment, to block other reviewers from approving flows that execute on sites not mapped to their identity. 

We rely on Globus Auth secrets to authenticate and authorize the remote execution of tasks. These secrets belong to a single user and can be used to authenticate to all sites to which that user has access. Therefore, it is recommended that a different GitHub Environment is created per user, each with one or more secrets.

To further enhance the security at the HPC sites, \gc{} endpoints have the ability to restrict what functions can be executed at a site. Thus, administrators may define the set of registered functions that may be executed. This ensures that the function itself has been pre-approved and deemed safe to execute on that endpoint.


\subsection{The CORRECT action}
\begin{figure}[t]
    \centering
    \includegraphics[width=\linewidth]{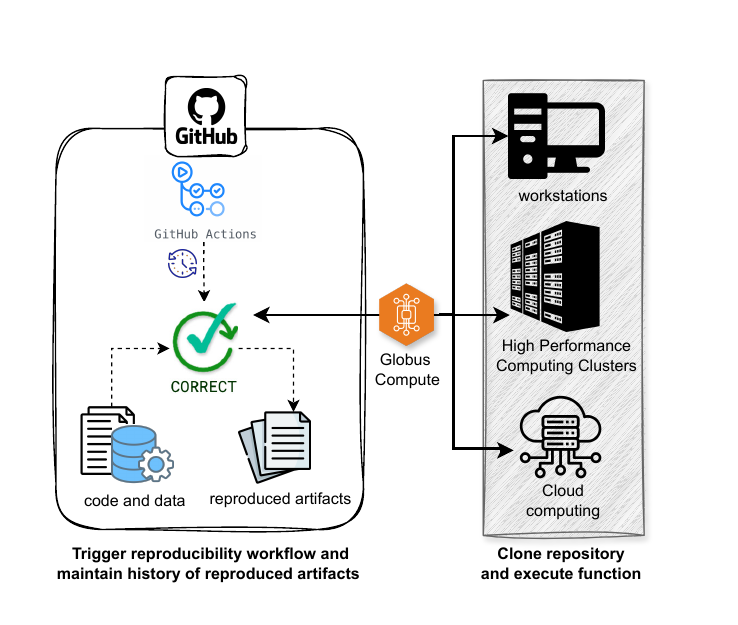}
    \caption{\name{} system overview.} 
    \label{fig:system-overview}
\end{figure}

\name{}, or COntinuous Reproducibility with a Remote Execution Computing Tool, 
is a GitHub Action that can be used to validate reproducibility across HPC and cloud resources. At its core, \name{} enables code defined on GitHub Actions to be remotely executed at arbitrary computing sites using Globus Compute (\autoref{fig:system-overview}). \name{} facilitates reproducibility by providing a public interface to trigger repeated runs, alter execution sites, and capture history of reproducibility evaluations. \name{} has been published to GitHub marketplace and is available here: \url{https://github.com/marketplace/actions/correct-action}.

Whereas HPC CI frameworks install runners directly on HPC infrastructure, \name{} runs within GitHub-hosted runners. By using this approach, access to the remote computing site is blocked without access appropriate authorization via defined secrets.
To communicate tasks to \gc{} Endpoints, the GitHub runners must first install Python and then invoke \name{}. \name{} will then communicate the user-provided inputs to the \gc{} service via its REST API and return the outputs once they are obtained.  

\autoref{fig:gcga-yaml} shows an example call to the \name{}. In this call, a \texttt{shell\_cmd} remotely executes the Python \texttt{tox} library. Other parameters to the action include a \gc{} client ID and secret as well as an endpoint UUID representing the \gc{} Endpoint on which the task should be executed. Should the remote function be a Python function, the function can be pre-registered with \gc{} and submitted to the action using the \texttt{function\_uuid} parameter. This feature also enables submission of only endpoint-approved functions to the \gc{} Endpoint.

When the action is invoked, it first verifies whether \gc{} exists on the runner, if it does not, it proceeds to pip install it. \name{} will then proceed to authenticate with Globus, obtaining a bearer token that will allow it to communicate with the \gc{} REST API. To ensure that the latest version of the code is being evaluated, \name{} will use a \gc{} function to clone the repository into a temporary directory. 
Once the repository is cloned, the user-specified function can be invoked. The standard output and standard error of the function execution is returned to the runner where it can be passed to subsequent functions or jobs within the workflow. The outputs can also be committed back to the repository or saved as action artifacts where they remain available for 90 days. Should either the cloning or user function execution fail, the workflow step will fail.

\name{} facilitates repeatability evaluations of scientific code by providing mechanisms to easily swap endpoints within the action workflow. Non-contributors can: 1) fork the repository, 2) instantiate their own \gc{} endpoints, 3) save their \gc{} secrets within a GitHub environment secret, 4) swap the endpoint UUIDs within the described workflow, and 5) trigger the workflow. Collaborators add their own infrastructure to the reproducibility evaluations by creating a new workflow file with their \gc{} endpoints, accessible through their specific environment.

\begin{figure}
    \centering
    \begin{minted}[
    gobble=4,
    frame=single,
    fontsize=\small,
    linenos,
  ]{yaml}
      - name: Run tox
        id: tox
        uses: globus-labs/correct@v1
        with:
          client_id: ${{ secrets.GLOBUS_ID }}
          client_secret: ${{ secrets.GLOBUS_SECRET }}
          endpoint_uuid: ${{ env.ENDPOINT_UUID }}
          shell_cmd: 'tox'
\end{minted}
\caption{Example Globus Compute Action step definition.}
    \label{fig:gcga-yaml}
\end{figure}




\section{Evaluation}

To evaluate the applicability of \name{} to HPC applications and software, we apply it to three different applications. These applications test various aspects of \name{}, from multi-site CI execution of a scientific application, logging and reporting of CI results for HPC software testing, and reproducibility of HPC research.


\subsection{Scientific Application Test Across Sites}

Protein docking aims to predict how two molecules bind to each other. It is commonly used in drug discovery to measure how well candidate drugs bind to their targets. As applications like protein docking would typically be used by a large community of users, we sought to evaluate how \name{} could be employed to evaluate reproducibility of the application across multiple sites. We use a Parsl-based implementation of protein docking (ParslDock)~\cite{raicu2023navigating} that uses machine learning to guide simulation and is deployed on HPC systems. 

For our experiment, we execute the ParslDock test suite at three different sites and record the duration of each test case using pytest. Our infrastructure test matrix includes a Chameleon Cloud~\cite{keahey2019chameleon} CHI@TACC Icelake instance and two ACCESS~\cite{ACCESS} HPC systems: TAMU FASTER and SDSC Expanse. On each system, we installed via Conda the Protein Docking application (\url{https://github.com/Parsl/parsl-docking-tutorial}) alongside AutoDock Vina (v1.2.6), VMD (v1.9.3), and MGLTools (v1.5.7). As neither FASTER or Expanse have outbound internet access on their compute nodes, we used \gc{} MEP functionality to define multiple providers at the sites. More specifically, we used a \texttt{LocalProvider} to first clone the repository on the login node. The cloned repository was saved to a compute-accessible location. To execute the test suite on compute nodes, we dynamically deploy a \gc{} user endpoint using the \texttt{SLURMProvider}.

\begin{figure}
    \includegraphics[width=\linewidth,trim=2.5mm 3mm 2mm 3mm,clip]{
        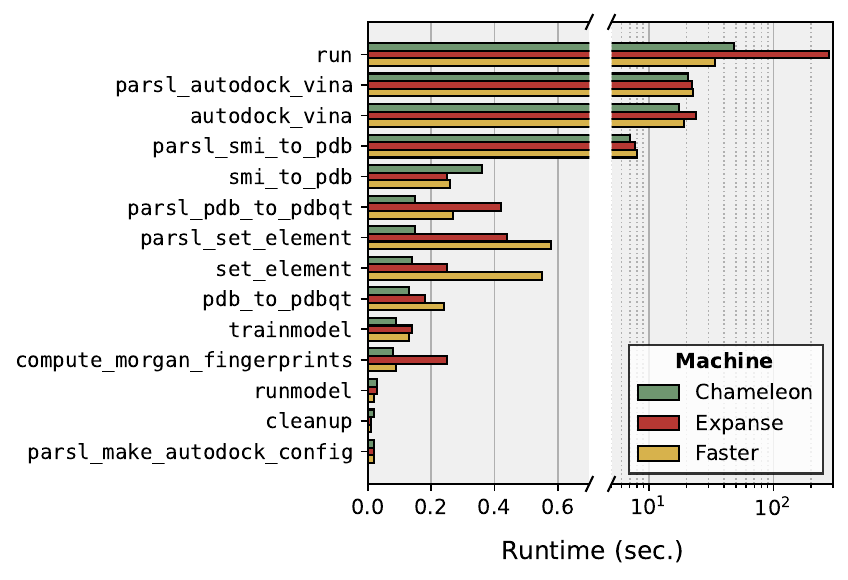
    }
    \caption{Runtimes of ParslDock tests on different machines.}
    \label{fig:runtimes}
\end{figure}

We evaluated \name{} functionality (i.e., feasibility and ease of use) on multi-site execution.
Our experiences running the ParslDock application showcase the ease by which \name{} can enable reproducibility across sites. It also highlights the ability to support diverse environments, including those with restricted network access to compute nodes.
\autoref{fig:runtimes} displays the results obtained from our benchmarks collected at the various sites.  We see that Chameleon outperforms other sites for most test cases. The short duration tests also highlight the benefits of adopting a FaaS based model. 

\subsection{HPC Software Testing}
PSI/J~\cite{hategan23psij} is a Python library designed to increase the portability of software---particularly workflow systems---across different HPC systems. 
PSI/J addresses the portability problem in HPC by providing an abstraction layer to the various available HPC schedulers.

Since PSI/J focuses on abstracting HPC schedulers, it is critical that it be regularly tested against deployed schedulers. While researchers have deployed schedulers inside containers, these do not broadly match the configurations of schedulers deployed on sites~\cite{hategan23psij}.
As a result, PSI/J must be tested directly on HPC sites, and its testing cannot be limited to a single site. PSI/J currently provides a mechanism for CI across different HPC that relies on cron jobs for automated, periodic execution of the test cases. The security relies on authenticated users deploying the cron job in their local accounts. 
However, the cron job is responsible for pulling the updated code regularly, which may pose some security risks. To mitigate potential security risks from automated execution of GitHub code, PSI/J provides three options for pulling code from the remote repository: 1) from the main branch, 2) stable and core branches, or 3) all branches part of a pull request that have been tagged by a core developer, ensuring that all code tested on an HPC machine has at least been reviewed by a core developer. However, it is not able to map a contributor or developer to a specific local account. PSI/J's cron job publishes test results back to the community via a public dashboard. 

In this experiment, we test the feasibility of expressing PSI/J CI jobs using \name{}. Our approach removes the need 
for cron jobs at individual sites. We evaluate our CI workflow by executing PSI/J CI tests on Purdue Anvil CPU. 
The metrics of our evaluation are the ability to define the same type of CI jobs as are already expressed by PSI/J and the ability to return similar data.

For our setup, we configure and instantiate a \gc{} MEP 
on the Anvil login node with PSI/J v0.9.9 installed within a Conda environment. The MEP is setup to use the \texttt{LocalProvider} since test cases must be run on the login node. On GitHub, the workflow is configured to run the action and extract the \texttt{stdout} and \texttt{stderr} generated and store them as artifacts---regardless of whether the tests pass or fail. 
Configuration of the test cases, if any, would be expressed within the \texttt{shell\_cmd} to \texttt{pytest} within the action.

\begin{figure}[!tbp]
    \centering
    \begin{minipage}[b]{\linewidth}
        \includegraphics[
            width=\textwidth,
            trim={0 5cm 0 0},
            clip
        ]{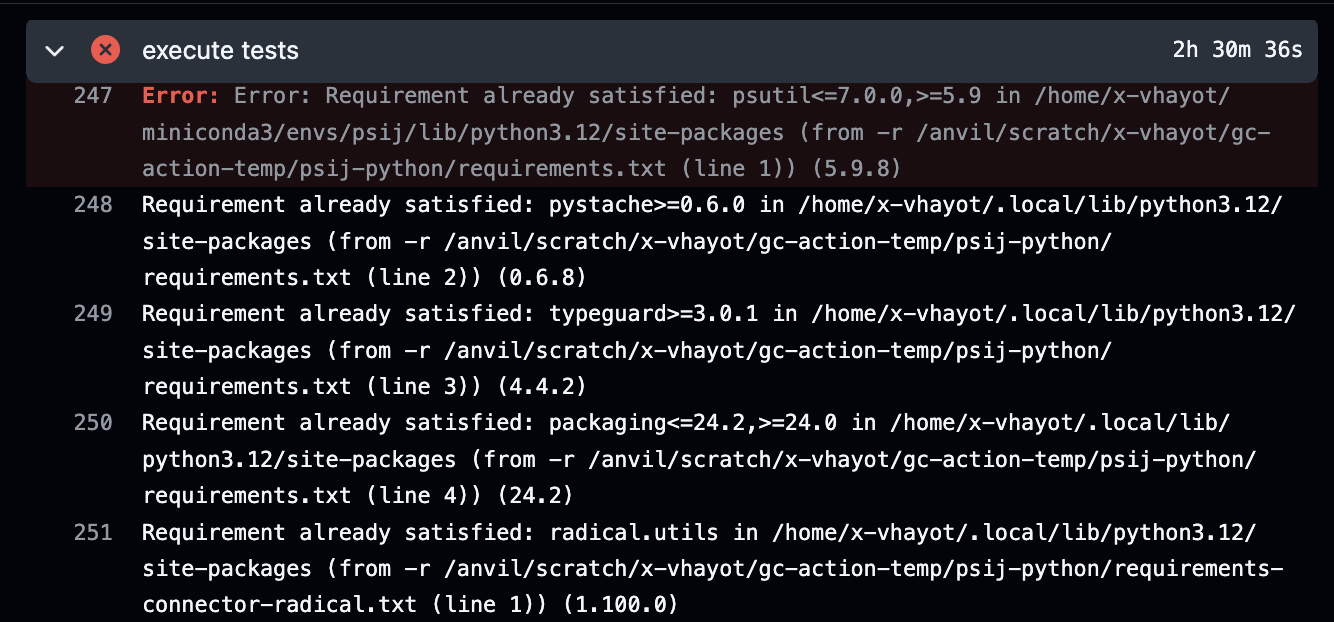}

    \end{minipage}
    \hfill
    \begin{minipage}[b]{\linewidth}

        \includegraphics[
            width=\textwidth,
            trim={0 1cm 1cm 0},
            clip
        ]{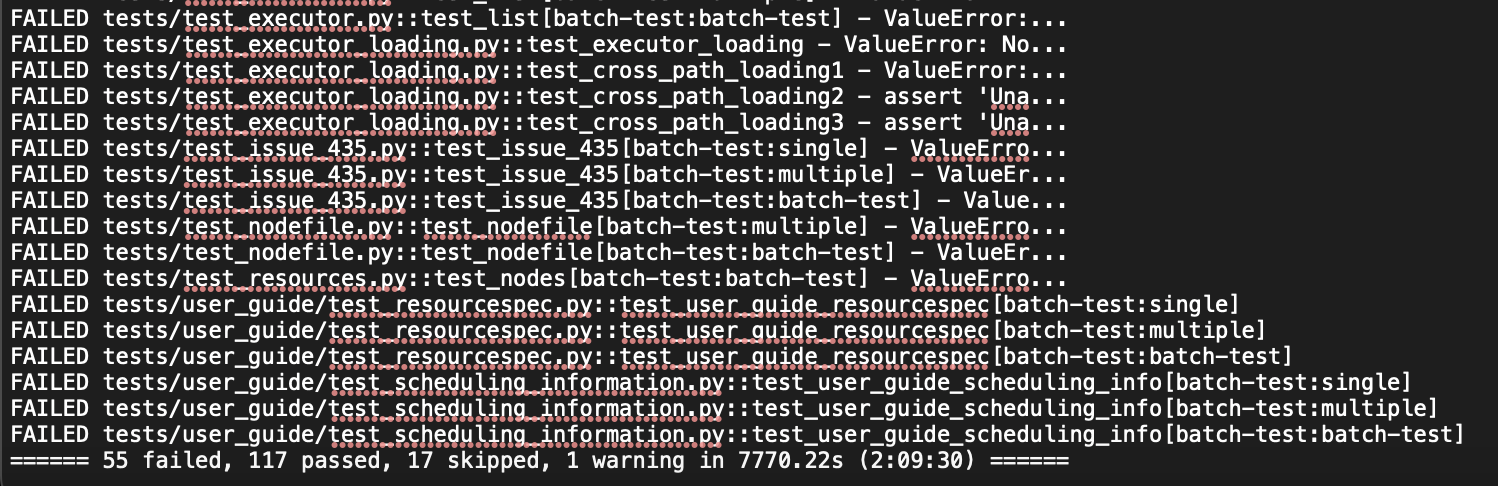}
        
    \end{minipage}
    \caption{PSI/J test invocation failure. \ul{Top}: Error and full execution \texttt{stdout} reported back to GitHub runner. \ul{Bottom}: Execution \texttt{stdout} stored within a GitHub Action artifact.}
    \label{fig:psij-fail}
\end{figure}

With \name{}, the PSI/J pytest runs did not succeed due to an error in the PSI/J codebase. \name{} reported the error by printing the failure directly in GitHub Action's UI interface and attaching the stdout/stderr outputs as artifacts to the CI run (\autoref{fig:psij-fail}). While the output stored here is not identical to the information reported on the PSI/J web UI, there is nothing that prohibits the same output from being generated and stored as both the PSI/J cron job and \name{} are essentially invoking a script via a shell command. In the case of \name{}, the command executed was the recommended \texttt{pytest} command, however, the CI job can be adapted to \name{} and include reporting back to the PSI/J dashboard.

\subsection{Reproducing an HPC Paper}

The KaMPIng library provides MPI bindings for C++~\cite{hespe2024kamping}. The KaMPIng paper was awarded the Best Reproducibility Advancement Award at SC 2024. Using the information detailed in their Artifacts Evaluation, we attempt to reproduce their results using \name{}. While most of the paper's results were said to be reproducible via downscaled experiments on a Chameleon Cloud VM, some required access to larger HPC (authors recommended SuperMUC-NG or HoreKa). 
It is in these instances that we believe the \name{} action to be useful in asserting reproducibility. With proven execution records (that can be reexecuted in an automated manner) and adequate provenance records (i.e., software environment setup), reproducibility reviewers can visualize the execution logs and evaluate reproducibility without re-execution of test cases. Furthermore, \name{} can easily facilitate execution of the test cases on  different target machines if desired by reviewers. 

\begin{sloppypar}
The KaMPIng artifacts are provided as a set of bash scripts that are designed to be executed within a Docker image published in the GitHub Container Registry (\url{https://github.com/kamping-site/kamping-reproducibility}). To execute these artifacts with \name{}, we created a GitHub workflow with each step representing the invocation of a single artifact. Execution outputs were all stored as Workflow Artifacts using the action \texttt{actions/upload-artifact@v4}. We executed these experiments on a Chameleon Cloud CHI@TACC IceLake instance. To ensure tasks were executed within the provided container, we configured and started a \gc{} MEP instance within the container; alternatively, the MEP could be configured to deploy the container dynamically. 
\end{sloppypar}

Our workflow execution demonstrates that all the Artifact Evaluation experiments pass with \name{}, with execution stdout and stderr published alongside the workflow execution. Using these published data alone, we compare our findings with those reported in the paper. 
We find that our execution with \name{} was able to reproduce the artifacts that were reported to be reproducible on Chameleon Cloud.

\section{Discussion}

While there is a growing demand for ensuring reproducibility of HPC experiments, achieving reproducibility in practice is challenging. 
Automation can address this gap by providing adequate accounting of the execution. However, automating experiments for reproducibility requires support to facilitate re-execution on other resources. 
Many HPC collaborative projects rely on CI approaches to test code functionality at different sites, using available HPC-specific CI solutions. Nevertheless, existing HPC CI solutions tend to be available only at single sites, as a result of site-specific security requirements. To address shortcomings of existing approaches, we developed a prototype, \name{}, for multi-site CI that can meet the basic security requirements of HPC sites. 
We describe here how \name{} addresses user needs with respect to ease of use, security, and overheads, and conclude with a discussion on the current limitations of our approach and potential future directions.  

\subsection{Ease of Use}

\name{} runs exclusively on the runner and requires only  
installation of Python
prior to executing the action. The action itself is simple to use, requiring only a few input parameters (i.e., client ID/secret, endpoint UUID, shell command or function UUID, and any arguments or configuration parameters). We chose GitHub Actions as a CI framework due to its ubiquity and collaborative functionalities; however, \name{} can be adapted for use with frameworks like GitLab CI/CD. 

For \name{} to deploy tasks across multiple computing sites, it requires that \gc{} Endpoints are configured to be used at each site. For our experiments, we relied on \gc{} MEPs 
as they can be reconfigured at execution time. For example, on sites like FASTER and Expanse, 
compute nodes are isolated from the internet, this meant we needed to leverage \gc{} MEP templates
to set different configurations when cloning the repository and executing the tests. However, standard \gc{} endpoints can also be used, as long as cloning and code execution can occur on the same nodes and do not require additional configuration. Users can reuse existing MEPs for sites that have them configured, eliminating the need for users to configure the endpoints themselves.

\subsection{Security}
\name{} relies on the security features provided by GitHub Actions and \gc{} to ensure authenticated and authorized resources use. \gc{} secrets are used to authenticate the running process with a Globus-authenticated user. \gc{} secrets are stored in GitHub as repository environment secrets  for use by the action. We recommend the use of GitHub environments with the requirement that the user who owns the client identity is the \textit{sole reviewer} of that action run. Although this may limit complete automation, it ensures that the user is always aware of the purpose of code executing with their credentials. While this may be problematic for nightly builds, basic test cases can be executed on cloud infrastructure, as long as they are part of a separate workflow, awaiting approval for execution on HPC.
Security can be further enhanced by restricting the functions that can be executed by the \gc{} endpoint. This can be specified within the endpoint configuration. It requires, however, that the functions are pre-registered with \gc{}. Any function that is not approved to execute on the endpoint will fail.

\subsection{Overhead on HPC infrastructure}

\name{} relies on \gc{} to manage the execution of tests. 
While single-user endpoints run a process per endpoint, MEPs can be configured to fork user endpoint processes for each user. Importantly, both single- and multi-user endpoints can be configured to dynamically provision compute resources, for example via a batch scheduler, to run tests.

With the use of GitHub environments for the management of workflow secrets, it is difficult to execute many tasks on the HPC infrastructure in rapid succession, thus preventing there from being too many tasks submitted to the batch scheduler at once. Furthermore,  \gc{} relies on a pilot job model and thus tasks can be executed on the pilot rather than requesting an allocation for each task.

\subsection{Limitations and Future Directions}

While \name{} demonstrates the ability to execute tasks at multiple remote sites and report results back to the GitHub web interface, there are several challenges that remain to be addressed. For example, since the tasks are invoked remotely from the runner, output files are not copied to the runner. External publishing mechanisms can be invoked from the \gc{} function to copy the outputs to a web accessible location, or be returned by the \gc{} function as output. While the latter is possible with Python functions (as long as their size does not exceed \gc{} limits), shell functions can only return standard output and error.

Another challenge is displaying the resource configuration at each invocation action. Without information about the environment, users can only see the results of previous executions, but that alone cannot validate reproducibility without access to the environment. Currently, we only demonstrate \name{'s} ability to invoke remote functions and return the results of their execution. A secondary call to \name{} could be made to capture a trace of the system's software environment and publish it as a workflow artifact. Alternatively, CORRECT could be extended to rely on containerization for task execution, requiring users to use published publicly-available containers within the \name{} workflows.

As GitHub artifacts remain available for only 90 days, it may be necessary to persist flow run executions to a more permanent location. GitHub provides commit actions that can commit files generated by previous steps to the repository. Alternatively, new steps could be added to the workflow to publish artifacts to external data repositories like Zenodo. 

For CI execution at HPC sites for which this action does not meet security requirements, it will be necessary to explore different methods for combining \name{} with approved CI frameworks. Furthermore, an open research challenge is to investigate methods which preserve automation while satisfying security requirements. 



\section{Summary}
While many initiatives aim to increase reproducibility of HPC research, most HPC research is not fully reproducible. HPC research outcomes are often tied directly to the infrastructure on which they were evaluated, and the lack of access to resources creates significant barriers for reproducibility. 
Continuous reproducibility, or automation through continuous integration, is a promising approach to ensure the 
reproducibility of scientific code; however, at present most approaches are limited to a single site. 
We proposed \name{}, a GitHub Action-based approach that uses \gc{} for remote code execution, as a means to address the basic security requirements of HPC while enabling automated multisite code execution within a single workflow. Using \name{} we demonstrated that we can easily reproduce results using an automated GitHub Action workflow. Future work will focus on improving security and accounting of reproducibility tests.

\begin{acks}
This material is based upon
work supported by the U.S. Department of Energy (DOE),
Office of Science, Office of Advanced Scientific Computing
Research, under Contract DE-AC02-06CH11357, and the National Science Foundation (NSF) under grant 2004894. Results presented in this paper were obtained using the Chameleon testbed supported by the National Science Foundation and used Expanse at SDSC, Anvil at Purdue University, and FASTER at Texas A{\&}M University, through allocation  CIS230030 from the Advanced Cyberinfrastructure Coordination Ecosystem: Services {\&} Support (ACCESS) program, which is supported by U.S. National Science Foundation grants \#2138259, \#2138286, \#2138307, \#2137603, and \#2138296.
\end{acks}

\bibliographystyle{ACM-Reference-Format}
\bibliography{bibliography}

\end{document}